\documentclass[11pt,onecolumn,journal]{IEEEtran}

\IEEEoverridecommandlockouts

\usepackage[T1]{fontenc}
\usepackage{amssymb,amsmath,epsfig}
\usepackage[latin5]{inputenc}
\usepackage{graphicx,times,latexsym}
\usepackage{psfrag,cite}
\usepackage{setspace}
\usepackage{algorithmic, algorithm}
\usepackage{times}

\begin{document}

\title{Factor Graph Based LMMSE Filtering for Colored Gaussian Processes}

\author{P\i nar~\c{S}en$^*$,~\IEEEmembership{Student~Member,~IEEE,}
Ali~\"{O}zg\"{u}r~Y\i lmaz,~\IEEEmembership{Member,~IEEE}
\thanks{The authors are with the Dept. of Electrical and Electronics Engineering, Middle East Technical University (psen@metu.edu.tr, aoyilmaz@metu.edu.tr), Ankara, Turkey. Parts of this work are accepted to 2014 IEEE Wireless Communications and Networking Conference (WCNC 2014), Istanbul, Turkey.}}%
\markboth{IEEE Signal Processing Letters}%
{Submitted paper}
\maketitle

\begin{abstract}
We propose a low complexity, graph based linear minimum mean square error (LMMSE) filter in which the non-white characteristics of a random process are taken into account. Our method corresponds to block LMMSE filtering, and has the advantage of complexity linearly increasing with the block length and the ease of incorporating the $\textit{a priori}$ information of the input signals whenever possible. The proposed method can be used with any random process with a known autocorrelation function with the help of an approximation to an autoregressive (AR) process. We show through extensive simulations that our method performs very close to the optimal block LMMSE filtering for Gaussian input signals.
\end{abstract}




%
\IEEEpeerreviewmaketitle

\IEEEpubidadjcol

\section{INTRODUCTION}
\label{sec:intro}
Together with the development of factor graphs and Gaussian message passing algorithms on linear state space models, filtering operations have lately been implemented with less computational complexity and less need for memory~\cite{Loeliger2004,Loeliger2007}. One example is the LMMSE filtering which is in fact equivalent to performing two-way Kalman filtering operations through a factor graph under Gaussian assumption~\cite{Loeliger2004}. The recently studied version of the LMMSE filtering was implemented on a factor graph under additive white Gaussian noise in~\cite{Guo2008,Loeliger2006} with a linearly increasing complexity over block size of the input signal whereas the computational complexity of the conventional block LMMSE filtering is increasing approximately with the cube of the block size. 

The factor graph approach to LMMSE filtering under white Gaussian noise in~\cite{Guo2008} provides a practical receiver structure particularly for the inter-symbol interference (ISI) channel which is frequently encountered in wireless communications. However, there are other problems in the literature in which the characteristics of non-white noise processes are needed to be taken into consideration. For example, in Faster than Nyquist (FTN) Signaling method~\cite{Anderson2012} and channel shortening for long, sparse ISI channels~\cite{Radosevic2011}, the inherent non-white noise processes are handled by various solutions including whitening filters. In addition, colored noise processes also appear in the radar problems\cite{Jameson2005} and the speech enhancement problems~\cite{Gibson1991}. 

Although forward Kalman filtering matrix operations are adapted for the noise statistics with Gaussian AR process in~\cite{Gibson1991}, there is no work on factor graphs which includes the effect of the colored noise in the literature within the knowledge of the authors. Hence, what we propose is a factor graph based LMMSE filtering approach which implements two-ways Kalman filtering operations through the graph with the ability of including the characteristics of the non-white Gaussian noise. We basically extend the state variables on the factor graph of~\cite{Guo2008} by joining them with the variables of the noise process as introduced in~\cite{Gibson1991}. The proposed method which can be generalized to other noise statistics through an approximation is studied under the Gaussian AR noise processes in this paper. Through extensive simulations it is shown that the proposed technique which has the advantage of linearly increasing computational complexity and less requirement of memory performs very close to the optimal block LMMSE filtering solution for Gaussian input signals. 

Another benefit of the proposed method comes from its factor graph based structure in which the existing $\textit{a priori}$ information of the input signals can be effortlessly incorporated as needed in many iterative communication receivers. Hence, it can be a practical way of block LMMSE filtering for the mentioned problems with non-white noise processes.  

The paper is organized as follows. The system model is described in Section~\ref{sec:system_model}. Section~\ref{sec:lmmse_graphs} presents a general graph based implementation of LMMSE filters and our proposed low complexity LMMSE filter for wide-sense stationary colored noise processes. In Section~\ref{sec:Simulation}, performance results of the proposed method are provided in comparison with the results of the block LMMSE filter and the LMMSE performed under a white noise assumption. Section~\ref{sec:conclsn} concludes the paper.

\section{SYSTEM MODEL}
\label{sec:system_model}

The notations used in the paper are as given below. Lower case letters (e.g., $x$) denote scalars, lower case bold letters (e.g., $\mathbf{x}$) denote vectors, upper case bold letters (e.g., $\mathbf{X}$) denote matrices. For a given random variable $x$; $m_x$, $v_x$ and $w_x$ denote its mean, variance and weight respectively where $w_x=v_x^{-1}$. For a given vector random variable $\mathbf{x}$; $\mathbf{R_x},\mathbf{m_x},\mathbf{V_x},\mathbf{W_x}$ denote 
its autocorrelation matrix, mean vector, covariance matrix and weight matrix respectively where $\mathbf{W_x}=\mathbf{V_x}^{-1}$. The indicators $()^*$ and $()^H$ denote conjugate and Hermitian transpose operations respectively. Finally, $E\{\}$ denotes the expectation operator.

\begin{figure}[htbp]
   \centering
   \includegraphics[width=0.45\textwidth]{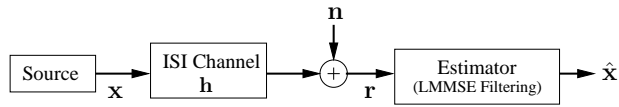}
   \caption{System Model}
   \label{fig:system_model}
\end{figure}

We consider the system given in Figure~\ref{fig:system_model}. 
The source sequence $\{x(k)\}$ is generated according to a zero-mean circularly symmetric complex white Gaussian process with a variance of $1$, i.e., $x(k) \sim CN(0,1)$ and the noise sequence $\{n(k)\}$ is generated according to a $p^{th}$ order AR process as defined in~\cite{Bhat1972}
\begin{align}
\label{eqn:ar_noise}
n(k) = \sum^p_{j=1}{a(j)n(k-j)} + w(k),
\end{align}
where $w(k)$ is a zero-mean circularly symmetric complex white Gaussian process with a variance of $\sigma^2_w$, i.e., $w(k) \sim CN(0,\sigma^2_w)$, and $a_j$'s are known AR process parameters. The multipath channel effect which is a commonly observed problem in wireless communications can also be included in the system model via an ISI channel gain vector of $\mathbf{h}$ with $L+1$ taps. Then, the noisy observation $r(k)$ at time $k$ is
\begin{align}
\label{eqn:observation}
r(k) = \sum_{i=0}^{L} h(i)x(k-i) + n(k) \;\; k=1,2,\ldots,N+L
\end{align}
where $N$ is the length of the input sequence, $E_s$ is defined as the average transmitted signal energy, i.e. $E_s \triangleq \sum_{i=0}^L E\{\vert h(i) \vert^2\}$ and $N_0$ is defined as the variance of the zero-mean noise process, i.e., $N_0 \triangleq E\{\vert n(k) \vert^2\}$. Parameters $N_0$, $a_j$'s and $\sigma^2_w$ are related by Yule-Walker equations in~\cite{Bhat1972}. We keep $N_0$ constant by adjusting $a_j$'s and $\sigma^2_w$ accordingly.

\section{GRAPH BASED LMMSE FILTERING}
\label{sec:lmmse_graphs}
\subsection{General Graph Approach to LMMSE Filtering}
\label{sec:LMMSE_compact}

\begin{figure}[htbp]
   \centering
   \includegraphics[width=0.47\textwidth]{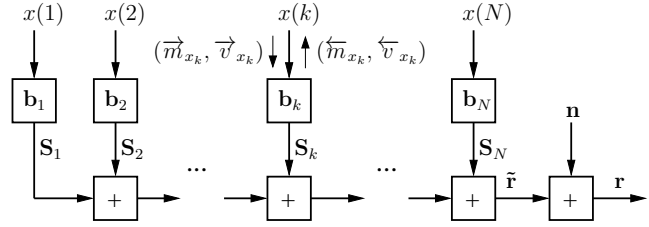}
   \caption{General Factor Graph of LMMSE Filtering Operation}
   \label{fig:compact_FG}
\end{figure}

A general factor graph based LMMSE filter for a system described via (\ref{eqn:observation}) was previously proposed in~\cite{Loeliger2007}. The factor graph which is applicable for any kind of stationary noise process is shown in Figure~\ref{fig:compact_FG} where $\mathbf{b}_k$ denotes the $k^{\textit{th}}$ column vector of the channel convolution matrix and $\mathbf{n}$ denotes the noise vector which are both length of $(N+L)$. It should be noted that each branch on the graph corresponds to either a set of state variables in vector form or a single state variable. For example, $\mathbf{S}_k$ represents the state variable vector of $(\mathbf{b}_k x(k))$, i.e., $\mathbf{S}_k = [0 \; \ldots \; 0 \; h(0)x(k) \; h(1)x(k) \; \ldots \; h(L)x(k) \; 0 \; \ldots \; 0]^T$, which is non-zero between $k^{\textit{th}}$ and $({k+L})^{\textit{th}}$ entries, and $\mathbf{n}$ represents the state variable vector of $[n(1)\;n(2)\ldots\;n(N+L)]^T$.

The block LMMSE filtering operation is implemented with the help of this factor graph by applying the Gaussian message passing rules in~\cite{Loeliger2007} which operates on the mean and variance values of the state variables. This message passing algorithm is performed through a forward recursion following the same directions as the arrows on the graph and a backward recursion following the reverse directions. The idea of using existing $\textit{a priori}$ information in the LMMSE filtering operation proposed in~\cite{Tuchler2002} is also included in the graph model by the incoming branches of $x(1),x(2),\ldots,x(N)$ and $\mathbf{n}$. The main purpose of the message passing algorithm is to calculate the $\textit{a posteriori}$ mean and variance values of the input source sequence by use of the incoming $(\overrightarrow{m}_{x_k},\overrightarrow{v}_{x_k})$ and outgoing messages $(\overleftarrow{m}_{x_k},\overleftarrow{v}_{x_k})$ resulting from the forward and backward recursions as below~\cite{Loeliger2006,Loeliger2007}
\begin{align}
\label{eqn:lmmse_var_apos}
{v}_{x_k}^{post}=&(\overrightarrow{v}_{x_k}^{-1}+\overleftarrow{v}_{x_k}^{-1})^{-1}, 
\end{align}
\begin{align}
\label{eqn:lmmse_mean_apos}
{m}_{x_k}^{post}=&{v}_{x_k}^{post}(\overrightarrow{v}_{x_k}^{-1} \overrightarrow{m}_{x_k}+\overleftarrow{v}_{x_k}^{-1} \overleftarrow{m}_{x_k}).
\end{align}

The incoming messages $\overrightarrow{m}_{x_k}$ and $\overrightarrow{v}_{x_k}$ are determined by the existing $\textit{a priori}$ information of the input signal and are set to $0$ and $1$ respectively in the setting studied here since the source is assumed to generate a white complex Gaussian signal sequence as $x(k) \sim CN(0,1)$. With the help of the observation vector $\mathbf{r}$, characteristics of the noise process and the incoming messages $(\overrightarrow{m}_{x_k}$,$\overrightarrow{v}_{x_k})$, the outgoing messages $\overleftarrow{m}_{x_k}$ and $\overleftarrow{v}_{x_k}$ can be reached through the message passing rules given in~\cite{Loeliger2007} as follows where $\tilde{\mathbf{W}}_{\mathbb{\kappa}}$ represents the auxiliary quantity for the specified vector $\mathbb{\kappa}$ as defined in~\cite{Loeliger2007}  

\begin{align}
\label{eqn:message_lmmse_var}
\tilde{\mathbf{W}}_{\mathbf{\tilde{r}}} \triangleq  & \left(\overrightarrow{\mathbf{V}}_{\mathbf{\tilde{r}}}+\overleftarrow{\mathbf{V}}_{\mathbf{\tilde{r}}}\right)^{-1}     \\
\tilde{\mathbf{W}}_{\mathbf{\tilde{r}}} =  & \tilde{\mathbf{W}}_{\mathbf{S}_k} \text{   for } k=1,2,\ldots ,N \\
\overleftarrow{v}_{x_k} = & {\tilde{w}_{{x}_k}}^{-1} - \overrightarrow{v}_{x_k} \\
=  & \left({\mathbf{b}_k^H \tilde{\mathbf{W}}_{\mathbf{\tilde{r}}}} \mathbf{b}_k \right)^{-1}-\overrightarrow{v}_{x_k}
\end{align} 
\begin{align}
\label{eqn:message_lmmse_mean}
\overleftarrow{m}_{x_k} = & {\tilde{w}_{{x}_k}}^{-1} \mathbf{b}_k^H \tilde{\mathbf{W}}_{\mathbf{\tilde{r}}} \overleftarrow{\mathbf{m}}_{\mathbf{S}_k}\\
=  & \left({\mathbf{b}_k^H \tilde{\mathbf{W}}_{\mathbf{\tilde{r}}}} \mathbf{b}_k \right)^{-1} \mathbf{b}_k^H \tilde{\mathbf{W}}_{\mathbf{\tilde{r}}} \overleftarrow{\mathbf{m}}_{\mathbf{S}_k}
\end{align} 
where
\begin{align}
\label{eqn:message_lmmse_mean_2}
\overleftarrow{\mathbf{m}}_{\mathbf{S}_k} = & \overleftarrow{\mathbf{m}}_{\tilde{\mathbf{r}}} - \sum_{l=1}^{N} {\mathbf{b}_l \overrightarrow{m}_{x_l}} + \mathbf{b}_k \overrightarrow{m}_{x_k}.
\end{align}

In our case, the noise is a zero-mean process with the autocorrelation matrix $\mathbf{R}_n$ which results in $\overleftarrow{\mathbf{V}}_{\mathbf{\tilde{r}}} = \mathbf{R_n}$ and $\overleftarrow{\mathbf{m}}_{\mathbf{\tilde{r}}} = \mathbf{r}$ in (\ref{eqn:message_lmmse_var}) and (\ref{eqn:message_lmmse_mean_2}) where ${\mathbf{\tilde{r}}}$ is the noise free observation vector. The computational complexity of this system is mainly determined by (\ref{eqn:message_lmmse_var}) and approximately $O(N^3)$ since the state variable vector $\mathbf{\tilde{r}}$ in (\ref{eqn:message_lmmse_var}) contains $N$ elements. We use the performance results of this algorithm as a benchmark to those of our proposed low complexity graph detailed in Section~\ref{sec:LMMSE_for_ar}.

\subsection{Enhanced Graph for LMMSE Filtering under AR Noise Process}
\label{sec:LMMSE_for_ar}
\begin{figure*}[t]
   \centering
   \includegraphics[width=0.72\textwidth]{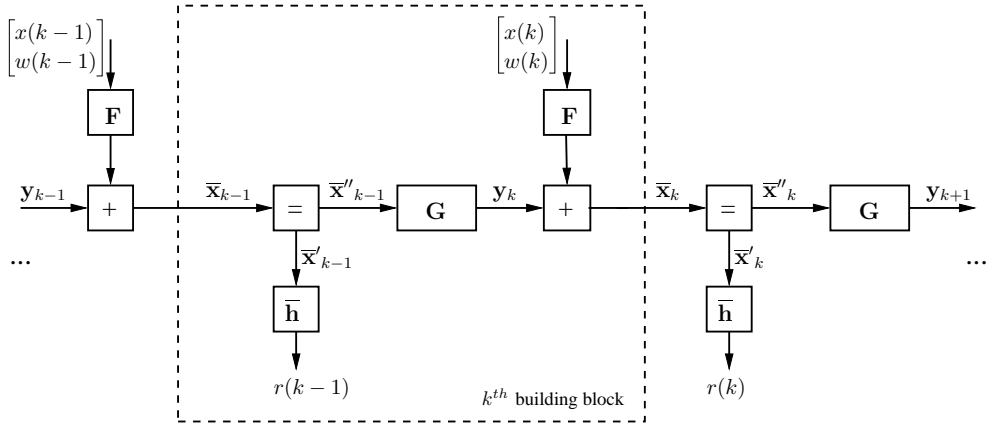}
   \caption{Enhanced Factor Graph for LMMSE Filtering under AR Noise Process}
   \label{fig:Gauss_Colored_LMMSE}
\end{figure*}

A later study in~\cite{Guo2008} accomplishes to decrease the $O(N^3)$ complexity of the graph based algorithm given in Section~\ref{sec:LMMSE_compact} to an $O(N(L+1)^2)$ complexity under the white noise scenario by separating the observation and noise vectors into scalar elements. However, this separation does not work under a non-white noise process since the observations conditioned on the input sequence are no longer independent. Hence, there is no work proposing a low complexity factor graph for colored noise case within the knowledge of the authors. On the other hand, the authors of \cite{Gibson1991} discuss the forward Kalman filtering operations under an order-$\textit{p}$ Gaussian AR noise process and propose to concatenate the set of state variables including the input sequence and the set of noise variables of length $p$. Hence, the main basis of our method is to apply this idea to the factor graph based LMMSE filtering operation studied in \cite{Guo2008} so as to implement the LMMSE approach having a linearly increasing computational complexity with $N$ under colored noise. In other words, we extend the state variable vector on the factor graph in~\cite{Guo2008} which is composed of only the input sequence by joining the noise variables and we make the necessary adjustments to preserve the smooth transition between the building blocks of the factor graph. This idea is also to be identified in Figure~\ref{fig:Gauss_Colored_LMMSE} later. 

In our factor graph representation the observation at time $k$ is the same as in (\ref{eqn:observation}). Hence, the $k^{\textit{th}}$ element of the observation vector $\mathbf{r}$ in (\ref{eqn:observation}) can be rewritten as
\begin{align}
\label{eqn:observation_ar_noise}
r(k) = \overline{\mathbf{h}} \: \overline{\mathbf{x}}_k, \; \text{where}
\end{align} 
\vskip -2em
\begin{align}
\label{eqn:state_h_vector}
\overline{\mathbf{h}} =& [h(L)\;\ldots\;h(1)\;h(0)\; 0 \; \ldots \; 0 \; 1]_{(1 \times L+p+1)}, \; \text{and} \\ 
\label{eqn:state_x_vector}
\overline{\mathbf{x}}_k =& [x(k-L)\;\ldots\;x(k-1)\;x(k)\; n(k-p+1) \; \ldots \; n(k)]^T.
\end{align}
We use (\ref{eqn:observation_ar_noise})-(\ref{eqn:state_x_vector}) in the construction of our factor graph representation which is depicted in Figure~\ref{fig:Gauss_Colored_LMMSE}. The joint state variables denoted by $\overline{\mathbf{x}}_k$ needs to be updated within each building block in the factor graph. For the transitions, we define
\begin{small}
\begin{align}
\label{eqn:FG_matrix_defn_G}
\mathbf{G} =& \left[ \begin{array}{cccc}
\mathbf{0}_{L \times 1} & \mathbf{I}_{L} & \mathbf{0}_{L \times 1} & \mathbf{0}_{L \times (p-1)} \\
0 & \mathbf{0}_{1 \times L} & 0 & \mathbf{0}_{1 \times (p-1)} \\
\mathbf{0}_{(p-1) \times 1} & \mathbf{0}_{(p-1) \times L} & \mathbf{0}_{(p-1) \times 1} & \mathbf{I}_{p-1} \\
0 & \mathbf{0}_{1 \times L} &  [\;\text{---------}\;\mathbf{a}\;\text{---------}\;] \hskip -5em &\\
 \end{array} \right], \\
\mathbf{F} =& \left[ \begin{array}{cc}
\mathbf{0}_{L \times 1} & \mathbf{0}_{L \times 1} \\
1 & 0 \\
\mathbf{0}_{(p-1) \times 1} & \mathbf{0}_{(p-1) \times 1}  \\
0 & 1 \\
 \end{array} \right]
\end{align} 
\end{small} 
where $\mathbf{a}$ is defined as the AR parameter vector, i.e., $\mathbf{a}= [a(p) \; a(p-1) \; \ldots \; a(1)]$, $\mathbf{I}_{j}$ denotes the identity matrix of size $j \times j$, and $\mathbf{0}$ denotes the all zero vector or matrix with specified sizes. The state variables are updated through the use of $\mathbf{F}$ and $\mathbf{G}$ as follows:
\begin{align}
\label{eqn:FG_state_defn_x_k1}
\overline{\mathbf{x}}_{k} &= \mathbf{F}\;[x(k) \quad w(k)]^T + \mathbf{y}_k, \; \text{where} \\
\label{eqn:FG_state_defn_y}
\mathbf{y}_k &= \mathbf{G}\; \mathbf{\overline{x}}_{k-1}.
\end{align}

The state space representation from (\ref{eqn:FG_state_defn_x_k1})-(\ref{eqn:FG_state_defn_y}) can be followed on the factor graph given in Figure~\ref{fig:Gauss_Colored_LMMSE}. In a similar fashion with Section~\ref{sec:LMMSE_compact}, the mean and variance values of the state vectors or scalars on the factor graph in Figure~\ref{fig:Gauss_Colored_LMMSE} are processed and updated by operating the Gaussian message passing rules derived in~\cite{Loeliger2007,Loeliger2006,Guo2008}. The main purpose of the factor graph of this state space model is to find the \textit{a posteriori} mean and variance values of the state variables $(\mathbf{m}_{\overline{\mathbf{x}}_k}^{post},\mathbf{V}_{\overline{\mathbf{x}}_k}^{post})$ by using the observation, noise characteristics and existing \textit{a priori} information of input signals through those operations. The Gaussian message passing rules are performed in both forward and backward directions resulting in the incoming $(\overrightarrow{\mathbf{m}}_{\overline{\mathbf{x}}_k},\overrightarrow{\mathbf{V}}_{\overline{\mathbf{x}}_k})$ and outgoing messages $(\overleftarrow{\mathbf{m}}_{\overline{\mathbf{x}}_k},\overleftarrow{\mathbf{V}}_{\overline{\mathbf{x}}_k})$ respectively. From these incoming and outgoing messages, \textit{a posteriori} mean and variance values of the state variables are obtained as follows~\cite{Guo2008,Loeliger2006}:
\begin{align}
\label{eqn:message_redilmmse_var_apos}
\mathbf{V}_{\overline{\mathbf{x}}_k}^{post}=&(\overrightarrow{\mathbf{V}}_{\overline{\mathbf{x}}_k}^{-1}+\overleftarrow{\mathbf{W}}_{\overline{\mathbf{x}}_k})^{-1}, 
\end{align}
\begin{align}
\label{eqn:message_redilmmse_mean_apos}
\mathbf{m}_{\overline{\mathbf{x}}_k}^{post}=&\mathbf{V}_{\overline{\mathbf{x}}_k}^{post}(\overrightarrow{\mathbf{V}}_{\overline{\mathbf{x}}_k}^{-1} \overrightarrow{\mathbf{m}}_{\overline{\mathbf{x}}_k} +\overleftarrow{\mathbf{W}}_{\overline{\mathbf{x}}_k} \overleftarrow{\mathbf{m}}_{\overline{\mathbf{x}}_k}).
\end{align}

We use these operations to reach the information on the state variables from which the source signal related part is extracted (noise part is stripped out) after performing the calculations given in (\ref{eqn:message_redilmmse_var_apos})-(\ref{eqn:message_redilmmse_mean_apos}). The \textit{a posteriori} mean and variance values of interest are indeed included in the first row of the joint state variables. It should be pointed out that the observation state variable $r(k)$ has a mean value of the $k^{\textit{th}}$ observation and variance value of $0$ which could be realized by a very small value, such as $10^{-5}$, through the message computations. Another important note is that the \textit{a priori} mean and variance values for the input signal at time $k$ ($m_{x(k)}^{prior}$,$v_{x(k)}^{prior}$) are involved in the graph in a way that they are concatenated with the mean and variance values of the zero mean white Gaussian noise $w(k)$ as in (\ref{eqn:FG_state_defn_x_k1}). Since $x(k)$ and $w(k)$ are independent random variables, the total \textit{a priori} information given to the $k^{\textit{th}}$ building block ($\mathbf{m}^{prior}_k$,$\mathbf{v}^{prior}_k$) can be written as:
\begin{align}
\label{eqn:red_lmmse_aprior_mean}
\mathbf{m}^{prior}_k =& \left[ \begin{array}{c}
m_{x(k)}^{prior}\\
0
\end{array} \right], 
\end{align}
\begin{align}
\label{eqn:red_lmmse_aprior_var}
\mathbf{v}^{prior}_k =& \left[ \begin{array}{cc}
v_{x(k)}^{prior} & 0\\
0 & \sigma^2_{w}
 \end{array} \right]. 
\end{align}
The source is assumed to generate a white complex Gaussian signal sequence as $x(k) \sim CN(0,1)$, so we can say that $m_{x(k)}^{prior}=0$ and $v_{x(k)}^{prior}=1$ in (\ref{eqn:red_lmmse_aprior_mean}) and (\ref{eqn:red_lmmse_aprior_var}).

The operations in each building block of the proposed factor graph includes matrix computations of size $(L+p+1)$ where $p$ is the number of AR parameters of noise process. Therefore, the complexity of our proposed algorithm is $O(N(L+p+1)^2)$ resulted from a similar way to~\cite{Guo2008}. It should be noted that our method provides a linearly increasing computational complexity with $N$ for a non-white Gaussian AR noise process. Besides, it is still possible to use the suggested approach for other non-white noise processes through an approximation as detailed in Section~\ref{sec:Yule-walker}.   

\subsection{Generalization to Other Processes}
\label{sec:Yule-walker}

Although we propose a new factor graph based LMMSE filter for a specific type of non-white noise process in Section~\ref{sec:LMMSE_for_ar}, it is still possible to use the proposed method for any kind of stationary noise process by means of an approximation. In other words, any wide-sense stationary noise process with known autocorrelation function can be approximated to an AR process by using the Yule-Walker equations~\cite{Bhat1972} given as 
\begin{equation}
\label{eqn:ar_corr}
R(j) = 
\begin{cases}
\sum_{i=1}^{p} a(i) R(-i) + \sigma_{w}^2 & \text{for } j=0  \\
\sum_{i=1}^{p} a(i) R(j-i) & \text{for }j>0.
\end{cases}
\end{equation}
where $R(i)$'s are the samples of the autocorrelation function. By choosing a proper value of $p$ and utilizing the first $p+1$ values of the autocorrelation function samples in (\ref{eqn:ar_corr}), approximate AR process parameters, $a(1),a(2),\ldots,a(p)$, and the variance of the additive white Gaussian noise term in (\ref{eqn:ar_noise}), $\sigma_w^2$, can be obtained. Although we have results related to this approximation~\cite{Pinar2014}, further elaboration is beyond the scope of this script. 

\section{SIMULATION RESULTS}
\label{sec:Simulation}

\begin{figure}[htbp]
   \centering
   \includegraphics[width=0.5\textwidth]{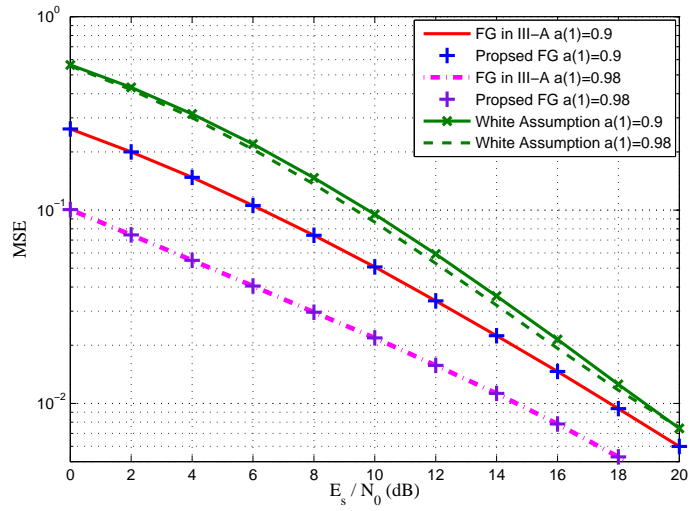}
   \caption{Performance of the Proposed Factor Graph Based LMMSE Filter under AR Gaussian Noise Process}
   \label{fig:MSE_all}
\end{figure}

In this section, performance results of the proposed LMMSE filtering method described in Section~\ref{sec:LMMSE_for_ar} for the system given in (\ref{eqn:observation}) with the input sequence length of $N=1000$ are presented in terms of mean square error (MSE) in Figure~\ref{fig:MSE_all}. For comparison, the performance results of the general graph based LMMSE filtering which corresponds to block LMMSE filter (optimal solution for our case) as mentioned in Section~\ref{sec:LMMSE_compact} and LMMSE filtering method studied in~\cite{Guo2008} under the assumption of a white Gaussian noise process are also given for the same configuration. The simulations are conducted for the noise processes of first order AR ($p=1$) with parameters $a(1)=0.9$ and $a(1)=0.98$ respectively for a multi-path static channel of $\sqrt{E_s/6}\:[1\;2\;0\;0\;0\;1]$. 
As verified from Figure~\ref{fig:MSE_all}, our proposed method is equivalent to the block LMMSE filter. In addition, it can be seen that the factor graph based LMMSE in~\cite{Guo2008} under the white noise assumption performs worse than the methods in which the non-white characteristics of the noise process are taken into consideration as expected. Moreover, we have observed that its performance loss increases as the noise correlation increases, i.e., $a(1)$ increases, since it does not use the correlation information of the noise process. Consequentially, although the performance loss of the white noise assumption may be ignored for the lower correlations of the noise process, our proposed factor graph seems to be a good choice with its low complexity and higher performance for applications involving high noise correlations in some communication and signal processing problems. Moreover, we have done other simulations for noise processes having $p>1$ under different number of channel taps which are not given here due to lack of space. However, the performance of the proposed graph always fits very well with the block LMMSE filtering solution.

\section{CONCLUSION}
\label{sec:conclsn}

In this paper, we propose a low complexity factor graph-based LMMSE filtering method for non-white noise processes which are encountered in some communication and signal processing problems such as FTN signaling, clutter suppression in radar systems and speech enhancement. Our method in which the statistics of the colored noise are taken into account seems to be an attractive solution to implement the LMMSE filtering operation owing to its linearly increasing computational complexity with the block length of the input signal. Although it is developed and simulated for the AR noise processes in this paper, the generalization to other stationary processes with known (or estimated) autocorrelation function through an approximation to a proper order AR process is feasible.

\bibliography{LMMSE_Filtering_Colored}

\end{document}